\documentclass[useAMS,usenatbib]{mn2e}
\usepackage{amsmath,array,amsfonts}
\usepackage{amssymb}
\usepackage{graphicx}
\usepackage{placeins}
\usepackage{float}
\usepackage{natbib}
\usepackage{pstricks}
\usepackage{verbatim}
\usepackage{subfigure}

\newcommand{\ltsima} {$\; \buildrel < \over \sim \;$}
\newcommand{\gtsima} {$\; \buildrel > \over \sim \;$}
\newcommand{\lta} {\lower.5ex\hbox{\ltsima}}
\newcommand{\gta} {\lower.5ex\hbox{\gtsima}}

\def\f{\frac}
\def\nn{\nonumber}
\def\ln{\mathrm{ln}}

\title{Combining Size and Shape in Weak Lensing}

\author[A. Heavens, J. Alsing, A. Jaffe]{Alan Heavens\thanks{e-mail:  a.heavens@imperial.ac.uk}, Justin Alsing, Andrew H. Jaffe \\
$^1$     Imperial Centre for Inference and Cosmology, Department of Physics, Imperial College, \\Blackett Laboratory, Prince Consort Road, London SW7 2AZ, U.K.\\}
\date{Accepted ;  Received ; in original form }

\begin{document}
\maketitle

\begin{abstract}
Weak lensing alters the size of images with a similar magnitude to the distortion due to shear.
Galaxy size probes the convergence field, and shape the shear field, both of which contain cosmological information. We show the gains expected in the Dark Energy Figure of Merit if galaxy size information is used in combination with galaxy shape.  In any normal analysis of cosmic shear, galaxy sizes are also studied, so this is extra statistical information comes for free and is currently unused.  There are two main results in this letter: firstly, we show that size measurement can be made uncorrelated with ellipticity measurement, thus allowing the full statistical gain from the combination, provided that $\sqrt{\rm Area}$ is used as a size indicator; secondly, as a proof of concept, we show that when the relevant modes are noise-dominated, as is the norm for lensing surveys, the gains are substantial, with improvements of about 68\% in the Figure of Merit expected when systematic errors are ignored.  An approximate treatment of such systematics such as intrinsic alignments and size-magnitude correlations respectively suggests that a much better improvement in the Dark Energy Figure of Merit of even a factor of $\sim 4$ may be achieved.
\end{abstract}
\section{Introduction} 

Weak gravitational lensing by the intervening nonuniform matter distribution has been recognised as a potentially very powerful tool for probing the growth rate of potential fluctuations and the geometry of the Universe through the distance-redshift relation.  Traditionally the statistic of choice has been cosmic shear --- the distortion in the shape of the image of a source \citep[see][and references therein]{Munshi2008}.  However, weak lensing has other effects, such as a magnification of the size of the image, and a corresponding change in the flux of sources.  In an ideal analysis, one would like to use all of this information.  Flux magnification is beginning to be explored (\citealt{VanWaerbeke2010,Hildebrandt2009,Hildebrandt2013}, Duncan et al.\ in prep) and after an early study \citep{Bartelmann1996} size magnification has begun to receive attention, both theoretically \citep{Casaponsa2012}, and observationally \citep{Schmidt2012}.  The latter study also considered magnitudes.  \cite{Casaponsa2012} showed that the convergence field can be recovered from the measured sizes of simulated galaxy images without any evidence of bias, provided the galaxies are larger than the point-spread function (PSF) and have S/N larger than 10.  These are very similar requirements for accurate estimation of shear, and since the shape measurement process also inevitably investigates size, this information comes for free.  The focus of this letter is two-fold: firstly to analyse what extra information is provided by size, and secondly to demonstrate that size and shape measurements can, with a careful definition of the size, be made uncorrelated, so we can use the full statistical power from adding size measurements.  It is intended to be the second step in a programme to develop more powerful combinations of weak lensing measurements to extract the full statistical power, and a number of questions are not addressed in this study, whose purpose is a proof-of-concept to illustrate that significant gains are possible.  With reasonable assumptions, we find that Figures of Merit for Dark Energy studies may be improved by significant factors, with no additional observational data required.

\section{Statistics of combined size and shear measurement}

In this section, we study what improvements in error bars we might expect from combining measurements of size and shape.  As the result is not quite as one might expect, we first illustrate the effect with a simplified case (a single tomographic bin and single mode), before performing Fisher matrix calculations to analyse the effect on a future survey designed to produce a large Dark Energy figure-of-merit.  We ignore systematic effects in this section, and consider them in Section~\ref{Results}.

Lensing effects are described by the transformation matrix mapping source angular positions to image positions,
\[ 
%\mathcal A(\vec{\theta})
\mathbfss{A}=\left( \begin{array}{c c}
1-\kappa-\gamma_1 & -\gamma_2 \\
-\gamma_2 & 1-\kappa+\gamma_1\\
\end{array} \right)\]
which defines the convergence field $\kappa$ and complex shear field $\gamma\equiv \gamma_1+i\gamma_2$. 
The magnification of surface area elements, $\nu$ is given by the determinant:
\begin{equation}
\nu = \frac{1}{\det \mathbfss{A}}=[(1-\kappa)^2-|\gamma|^2]^{-1}.\label{mag}
\end{equation}
If $|\kappa|$ and $|\gamma|\ll 1$ (which we assume throughout) this can be approximated by $\nu\simeq 1+2\kappa$,
so a length scale defined by the square root of the area, which we will see is a very useful definition of size, will scale to linear order by $1+\kappa$.

In the \citet{Limber1954} approximation, the angular power spectrum of the lensing potential between tomographic
redshift bins $i$ and $j$ is given by 
%\citep{Kaiser1992, Kaiser1998, Hu1999, Hu2002a, Takada2004}
\citep{Takada2004}
\begin{align}
C^{\phi\phi}_{\ell(i,j)} &= \f{4}{\ell^4}\left(\f{3\Omega_\mathrm{m}H_0^2}{2}\right)^2\int d\chi\;w_{(i)}(\chi)w_{(j)}(\chi)\chi_\mathrm{m}(\chi)^{-2} \nn \\
&\;\;\;\;\;\;\;\;\;\;\;\;\;\;\;\;\;\times(1+z(\chi))^2P\left(k = \f{\ell}{\chi_\mathrm{m}(\chi)}; \chi\right),
\end{align}
where $\chi$ is comoving distance, $P(k; \chi)$ is the 3D matter power spectrum and $\chi_\mathrm{m}(\chi)$ is the transverse comoving distance corresponding to comoving distance 
$\chi$. The lensing weight functions $w_{(i)}(\chi)$ are given by
\begin{align}
w_{(i)}(\chi)=
\begin{cases}
\f{\chi_\mathrm{m}(\chi)}{\bar{n}_i}\int_{\mathrm{max}(\chi_i,\chi)}^{\chi_{i+1}} d\chi\;p(\chi)\f{\chi_\mathrm{m}(\chi'-\chi)}{\chi_\mathrm{m}(\chi')}, & \chi \leq \chi_{i+1} \\
0, & \chi > \chi_{i+1}
\end{cases}
\end{align}
where $p(\chi)d\chi = n(z)dz$ for a galaxy redshift distribution $n(z)$, $\bar{n}_i$ is the number density in the $i$th bin, $\chi_i$ and $\chi_{i+1}$ are the boundaries of the $i$th tomographic bin. 
On the full 2D sky, the spherical harmonic expansion coefficients of the shear and convergence fields (associated with a particular tomographic bin) are related to those of
the lensing potential \citep[e.g.,][]{Hu2000,Castro2005}:
\begin{align}
\kappa_{\ell m(i)} &= -\f{1}{2}\ell(\ell+1)\phi_{\ell m(i)}\approx -\f{1}{2}\ell^2\phi_{\ell m(i)}, \nn \\
\gamma_{1,\;\ell m(i)} &= \f{1}{2}\sqrt{\f{(\ell + 2)!}{(\ell - 2)!}}\phi_{\ell m(i)}\approx \f{1}{2}\ell^2\phi_{\ell m(i)}, \nn \\
\gamma_{2,\;\ell m(i)} &= -\f{i}{2}\sqrt{\f{(\ell + 2)!}{(\ell - 2)!}}\phi_{\ell m(i)}\approx -\f{i}{2}\ell^2\phi_{\ell m(i)}.
\end{align}
Taking as the estimators $\hat{\kappa} = \ln(\lambda/\bar\lambda)=\kappa + \ln(\lambda_s/\bar \lambda)$ and $\hat{\gamma} = e = \gamma + e_s$, where $\bar\lambda$ is the mean size at the 
appropriate redshift and $\lambda_s$ is the unmagnified source size, the (cross) power spectra are given by 
%\citep{Kaiser1992, Kaiser1998, Hu1999, Hu2002a, Castro2005}
(e.g. \cite{Hu2002a})
\begin{align}
\hat{C}^{\kappa\kappa}_{\ell(i,j)} &= \f{1}{4}\ell^4C^{\phi\phi}_{\ell(i,j)} + \delta_{ij}\sigma_{\ln\lambda}^2/\bar{n}_i \nn \\
\hat{C}^{\gamma_1\gamma_1}_{\ell(i,j)} &= \hat{C}^{\gamma_2\gamma_2}_{\ell(i,j)} = \f{1}{4}\ell^4C^{\phi\phi}_{\ell(i,j)} + \delta_{ij}\sigma_{e}^2/\bar{n}_i \nn \\
\hat{C}^{\gamma_2\gamma_1}_{\ell(i,j)} &= \f{1}{4}\ell^4C^{\phi\phi}_{\ell(i,j)} \nn \\
\hat{C}^{\gamma_1\kappa}_{\ell(i,j)} &= \hat{C}^{\gamma_2\kappa}_{\ell(i,j)} = \f{1}{4}\ell^4C^{\phi\phi}_{\ell(i,j)},
\end{align}
where $\sigma_e$ and $\sigma_{\ln\lambda}$ are the dispersions in the intrinsic (complex) ellipticity and log-size $\ln\lambda_s$ respectively.

Since $\kappa_{\ell m},\;\gamma_{1,\;\ell m}$ and $\gamma_{2,\;\ell m}$ are complex, care must be taken in constructing the covariance matrix to ensure that
all of the information has been included correctly. Here we take our data vector to contain entries for the expansion coefficients and their complex conjugates,
i.e., $\mathbfit{d}^{(\kappa,\gamma)\mathrm{T}} = (\mathbfit{z}^{(\kappa,\gamma)\mathrm{T}}, \mathbfit{z}^{(\kappa,\gamma)*\mathrm{T}})$
for the combined shear-convergence data and $\mathbfit{d}^{(\gamma)\mathrm{T}} = (\mathbfit{z}^{(\gamma)\mathrm{T}}, \mathbfit{z}^{(\gamma)*\mathrm{T}})$
for the shear only case, where $\mathbfit{z}^{(\kappa,\gamma)} = (\dots\kappa_{\ell m (i)},\gamma_{1,\;\ell m(i)},\gamma_{2,\;\ell m(i)}\dots)^\mathrm{T}$ and
$\mathbfit{z}^{(\gamma)} = (\dots\gamma_{1,\;\ell m(i)},\gamma_{2,\;\ell m(i)}\dots)^\mathrm{T}$ contain the full set of relevant complex coefficients. Note that populating the data vector
with the real and imaginary parts of the relevant fields explicitly is entirely equivalent \citep[see, e.g.,][]{Picinbono1996}. Furthermore, to avoid duplication of information
only $m\geq0$ modes are included and care must be taken not to double count the $m=0$ modes for which $\kappa_{\ell0},\;\gamma_{1,\ell0}$ and $\gamma_{2,\ell0}$ are real
(recall that since $\phi$ is a real field, $\phi_{\ell m}\propto\phi^*_{\ell -m}$ and $\phi_{\ell0}\in\mathbb{R}$).
The full covariance matrix $\boldsymbol{\Gamma}$ of the data is defined as:
\begin{align}
\boldsymbol{\Gamma} = \langle \mathbfit{d} \mathbfit{d}^\dagger\rangle &= \left( \begin{array}{cc}
\langle \mathbfit{z}\mathbfit{z}^\dagger \rangle & \langle \mathbfit{z}\mathbfit{z} \rangle \\
\langle \mathbfit{z}\mathbfit{z} \rangle^* & \langle \mathbfit{z}\mathbfit{z}^\dagger \rangle^*
\end{array}\right) \nn \\
& = \left( \begin{array}{cc}
\mathbfss{C} & 0 \\
0 & \mathbfss{C}
\end{array}\right),
\end{align}
where in the second line we have used the fact that $\langle \mathbfit{z}\mathbfit{z} \rangle =0$ and $\mathbfss{C} = \langle \mathbfit{z}\mathbfit{z}^\dagger \rangle\in\mathbb{R}$.
Since different $\ell$ and 
$m$ modes are un-correlated for an all-sky survey (we relax this later), $\mathbfss{C}$ will be block diagonal with each $(\ell,m)$-mode contributing one diagonal block:
\begin{align}
\mathbfss{C}_\ell = \mathbfss{P}_\ell\otimes \mathbfss{X}_\ell + \mathbfit{n}^{-1}\otimes\mathbfss{N}_\sigma,
\label{Cov}
\end{align}
where
\begin{align}
\mathbfit{n} = \mathrm{diag}(\bar{n}_1,\;\bar{n}_2\dots),\;\;\mathbfss{P}_{\ell,ij} &= C^{\phi\phi}_{\ell(i,j)}, \nn
\end{align}
\begin{align}
\mathbfss{N}^{(\gamma)}_\sigma = \mathrm{diag}\left(\sigma_e^2, \sigma_e^2\right),\;\;\mathbfss{N}^{(\kappa,\gamma)}_\sigma = \mathrm{diag}\left(\sigma_{\ln\lambda}^2, \sigma_e^2, \sigma_e^2\right), \nn
\end{align}
\begin{align}
\mathbfss{X}^{(\gamma)}_\ell = \f{\ell^4}{4}\left( \begin{array}{cc}
1 & 1 \\
1 & 1 \end{array} \right),\;\;\mathbfss{X}^{(\kappa,\gamma)}_\ell = \f{\ell^4}{4}\left( \begin{array}{ccc}
1 & 1 & 1 \\
1 & 1 & 1 \\
1 & 1 & 1 \end{array} \right)
\end{align}
and $\otimes$ is the tensor product.

The Fisher matrix, $\mathbfss{F}_{\alpha\beta}$,  is the negative expectation of the second derivative of the log-likelihood with respect to the model parameters labelled by $\alpha$ and $\beta$. If the data can be assumed to be Gaussian distributed with fixed means, such that the covariance matrix is determined by the parameters of interest,
the Fisher matrix can be computed from the covariance matrix and its derivatives \citep{Tegmark1997}:
\begin{eqnarray}
\mathbfss{F}_{\alpha\beta} &=& \f{1}{2}\mathrm{Tr}\left[\mathbf{\boldsymbol{\Gamma}}^{-1}\boldsymbol{\Gamma}_{,\;\alpha}\boldsymbol{\Gamma}^{-1}\boldsymbol{\Gamma}_{,\;\beta}\right], \nn \\
&= &\mathrm{Tr}\left[\mathbfss{C}^{-1}\mathbfss{C}_{,\;\alpha}\mathbfss{C}^{-1}\mathbfss{C}_{,\;\beta}\right]\;,
\label{FisherEq}
\end{eqnarray}
where a subscripted comma refers to derivatives with respect to the following parameter.
Since $\mathbfss{C}$ is block diagonal, with each $(\ell, m)$ mode contributing a block $\mathbfss{C}_\ell$, the Fisher matrix can be written as a sum over modes
\begin{align}
\mathbfss{F}_{\alpha\beta} = f_\mathrm{sky} \sum_{\ell_\mathrm{min}}^{\ell_\mathrm{max}}\left(\ell + \f{1}{2}\right){\mathrm{Tr}}\left[ \mathbfss{C}_\ell^{-1}\mathbfss{C}_{\ell,\;\alpha}\mathbfss{C}_\ell^{-1}\mathbfss{C}_{\ell,\;\beta}\right],
\end{align}
where we have also included a factor $f_\mathrm{sky}$ to approximately account for incomplete sky coverage.

To illustrate, let us consider estimating the amplitude of the lensing potential power spectrum $C_\ell$ from a single mode and a single tomographic bin (so we drop the (i) subscript), ignoring for now systematic effects in shape and size, and assuming the same number density for both size and ellipticity.  In this case, for a shape and size analysis, the covariance matrix $\mathbfss{C}$, Eq.~(\ref{Cov}) is
\begin{align}
\mathbfss{C}^{(\kappa,\gamma)} = \left( \begin{array}{ccc}
c_\ell+\sigma_e^2/\bar n & c_\ell & c_\ell\\
c_\ell & c_\ell+\sigma_e^2/\bar n & c_\ell\\
c_\ell & c_\ell & c_\ell + \sigma^2_{\ln\lambda}/\bar n\end{array} \right).
\end{align}
where $c_\ell \equiv \ell^4 C_\ell/4$.  For a shape-only analysis, $\mathbfss C$ is the top-left $2\times 2$ sub matrix.

The Fisher matrix  given by Eq.~(\ref{FisherEq}) is a scalar in these cases, and reduces to (multiplying by 2 since the covariance matrix has two blocks of $\mathbf C$),
\begin{equation}
F^{(\kappa,\gamma)} = \frac{\bar n^2 \left(\sigma _e^2+2 \sigma _{\ln\lambda}^2\right){}^2}{\left[\sigma _e^2 \left(c_\ell \bar n+\sigma _{\ln\lambda}^2\right)+2 c_\ell \bar n \sigma _{\ln\lambda}^2\right]{}^2}
\end{equation}
and
\[
F^{(\gamma)}=\frac{4\bar n^2}{\left(2 \bar n c_\ell+\sigma_e^2\right)^{2}}.
\]
Therefore the error bar on $c_\ell$ is reduced by a factor 
\begin{equation}
\frac{\sigma_{\rm shear}}{\sigma_{\rm shear\ +\ size}} = 
\frac{(2R + 1) (2S + 1 )}{2 (2SR + S + R )}
\end{equation}
where $R\equiv \sigma_{\ln\lambda}^2/\sigma_e^2$ and $S\equiv \bar n c_\ell/\sigma^2_{e}$ is a measure of signal-to-noise.

We see that in the high S/N limit, there is no gain; essentially both size and shape are measuring the same quantity with a vanishingly small error bar.  Since the signal we are using here is the variance around the zero mean, there is no benefit.  The other limit is interesting; in the low S/N regime, the gain is a factor of 1.5 if $\sigma_e=\sigma_{\ln\lambda}$. i.e.  we estimate the variance with an error smaller by a factor $3/2$.

\section{Uncorrelated area and shape measurement}

We have so far assumed that the estimates of the shear and convergence are uncorrelated.  It is not obvious that this case be achieved, but in this section we demonstrate that for galaxies which have exponential brightness profiles, the estimate of $\sqrt{\rm area}$ is uncorrelated with the estimate of ellipticity when estimated using model-fitting methods such as {\em lens}fit \citep{Miller2007}.  For more complex morphologies, we would expect the correlations to be non-zero, but there is a reasonable expectation that they would be small.

We model the galaxy surface brightness $\mu$ with a thin, intrinsically circular, disk with an exponential profile with scale length $R$.  The apparent shape of the galaxy image is determined by the angle $\eta$ between the disk normal and the line of sight.  If the projected elliptical image has a position angle $\phi$, then after some algebra, the surface brightness may be written, as a function of polar coordinates ($r,\psi$),
\begin{equation}
\mu(r,\psi|\lambda,\epsilon,\phi,\mu_0)=\mu_0 \exp\left[-\frac{r\sqrt{1+e^2-2e \cos 2(\psi-\phi)}}{\lambda \sqrt{1-e^2}}\right]
\end{equation}
where $\mu_0$ is the central surface brightness,
\[
\lambda \equiv a \sqrt{\frac{1-e}{1+e}} 
\]
and $a, b$ are the semi-major and semi-minor axes, $e$ is the (magnitude of the) ellipticity, defined by $e=(a-b)/(a+b)$.  The area of the ellipse is $\pi a b = \pi \lambda^2$, so $\lambda$ is a measure of the square root of the area.  We ignore errors in the centroid in what follows.

If we estimate the four parameters of the model, $\bmath\theta = (\lambda,\epsilon,\mu_0,\phi)$ from a set of pixels with gaussian white noise errors, then we can obtain the covariance of the estimates from the Fisher matrix, here with the covariance matrix fixed \citep[e.g.,][]{Tegmark1997}:
\[
\mathbfss{F}_{\alpha\beta} = \frac{1}{2}{\rm Tr}\left[\mathbfss{C}^{-1}\frac{\partial \bmath{\mu}}{\partial \theta_\alpha}\frac{{\partial \bmath\mu}^T}{\partial \theta_\beta}+\mathbfss{C}^{-1}\frac{\partial \bmath\mu}{\partial \theta_\beta}\frac{{\partial \bmath\mu}^T}{\partial \theta_\alpha}
\right]\label{Fisher}
\]
where the covariance matrix $\mathbfss{C}$ is diagonal and proportional to the identity.  $\bmath\mu$ is a vector of the expected pixel values, having integrated the model over the pixel area and accounting for PSF effects (which we ignore in this analysis).  For this study, we replace the matrix summation by a continuum approximation and integrate over the image.  After some Mathematica algebra, the Fisher matrix simplifies to 
\begin{equation}
\mathbfss{F} \propto \left(
\begin{array}{cccc}
 \frac{3 \pi }{4} & 0 & \frac{\pi }{2} & 0 \\
 0 & \frac{3 \pi }{8 \left(1-\epsilon ^2\right)^2} & 0 & 0 \\
 \frac{\pi }{2} & 0 & \frac{\pi }{2} & 0 \\
 0 & 0 & 0 & \frac{3 \pi  \epsilon ^2}{2 \left(1-\epsilon ^2\right)^2} \\
\end{array}
\right).
\end{equation}
where the rows and columns correspond to the order $\bmath\theta = (\lambda,\epsilon,\mu_0,\phi)$.

The correlation matrix of the parameters is formed from the inverse of the Fisher matrix, and is
\begin{equation}
\mathbfss{F}^{-1} \propto \left(
\begin{array}{cccc}
 1 & 0 & -\sqrt{\frac{2}{3}} & 0 \\
 0 & 1 & 0 & 0 \\
 -\sqrt{\frac{2}{3}} & 0 & 1 & 0 \\
 0 & 0 & 0 & 1 \\
\end{array}
\right).
\end{equation}
From this we see that our estimate of $\lambda$ is completely uncorrelated with the position angle $\phi$, and strongly anti-correlated with the central surfae brightness $\mu_0$, as one might expect.  The important result is that there is no correlation with the ellipticity $e$, and this is not necessarily expected.   This arises from our choice of size parameter as $\lambda = \sqrt{{\rm Area}/\pi}$.  The choice of the semi-major axis as the size parameter in \cite{Casaponsa2012} is not nearly as useful, as it is highly correlated with ellipticity.

There are many effects which are not considered in this analysis, such as the PSF, pixelisation, a range of profiles and centroid errors, all of which may lead to some correlations between size and shape, but we expect on the basis of this calculation that these correlations will be small provided that $\sqrt{{\rm Area}}$ is used for size, and we will ignore them in this paper.

\section{Results}\label{Results}

\begin{figure*}
\includegraphics[width = 16cm]{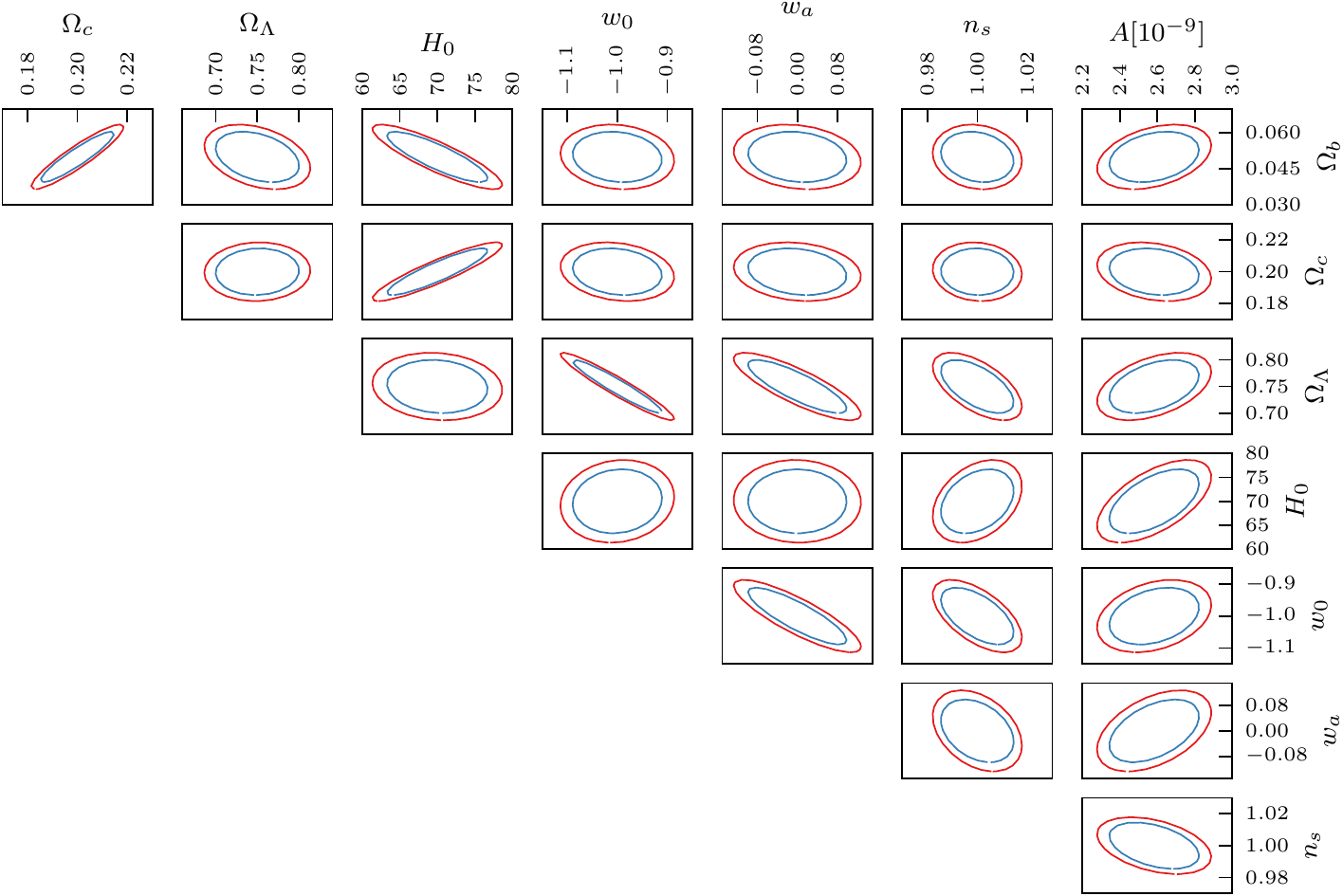}
\caption{This figure shows the marginal error plots from a Fisher matrix analysis of a Euclid-like weak lensing survey.  The Dark Energy FoM improves by 68\%. For details of survey parameters and assumptions, see text.}\label{fig:Blob}
\end{figure*}

\begin{figure}
\centering
\resizebox{8cm}{!}{\includegraphics{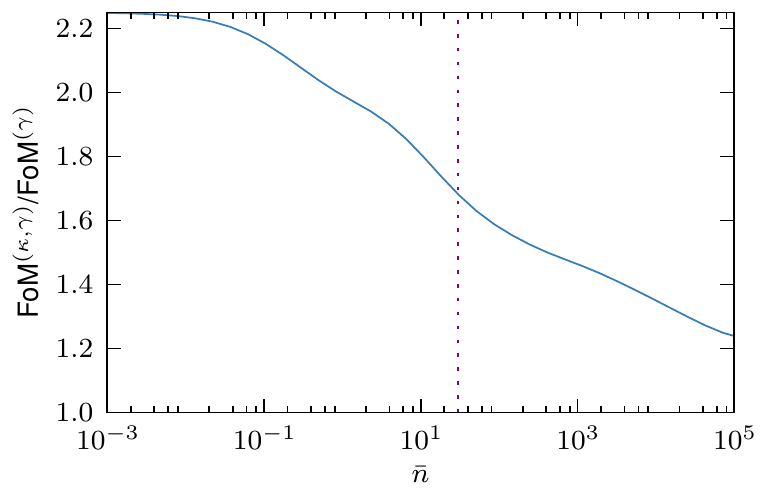}} 
\caption{The relative improvement in the Dark Energy FoM from shape plus size vs shape alone, as a function of the number density of galaxies, with fixed median redshift $z_m=0.9$.  A Euclid-like survey with $\bar n=30$ per square arcminute is shown by the vertical dashed line.}
\label{fig:FoM}
\end{figure}

We consider a 15000 square degree survey similar to that proposed for the ESA Euclid mission.  We assume a redshift distribution $n(z)\propto z^2 \exp[-(1.41 z/z_m)^{1.5}]$, with a median redshift $z_m=0.9$ and a mean number density $\bar n=30$ per square arcminute. We assume a dispersion in $\ln\lambda$ of $0.3$ \citep{Shen2003,Ferguson2004}, and $\sigma_e=0.3$ . We consider tomography with 10 bins between redshifts 0 and 2, with equal numbers per bin. We compute the lensing potential power spectrum for each bin using CAMB to compute the matter power spectrum, and vary the following cosmological parameters: $\Omega_b$, $\Omega_c$, $\Omega_\Lambda$, $h$, $w_0$, $w_a$, $n_s$, $10^9 A$, being, respectively, the density parameters in baryons, Cold Dark Matter and Dark Energy, the Hubble parameter in units of $100$km\,s$^{-1}$\,Mpc$^{-1}$, the Dark Energy equation of state parameters ($p/\rho = w_0 + w_a(1-a)$, where $a$ is the scale factor), the scalar spectral index, and the amplitude of fluctuations.  We have not included a number of effects, such as intrinsic alignments or photometric redshift errors in this proof-of-concept study, but the main interest here is the relative change in the errors when size magnification is added, rather than in absolute values. We show this two ways: firstly by showing the marginal errors of pairs of parameters, in Fig.~\ref{fig:Blob}, and secondly by computing the Figure of Merit (FoM) for Dark Energy, defined to be the inverse of the area$/\pi$ of the 1$\sigma$ contours of the expected likelihood in the $w_0,w_a$ plane, marginalised over all other parameters.  This is shown as a function of  $\bar n$ in Fig.~\ref{fig:FoM}.  For $\bar n=30$ the FoM is increased from 293 to 492, an improvement of 68\%.

\subsection{Systematic Errors}

\begin{figure}
\centering
\resizebox{8cm}{!}{\includegraphics{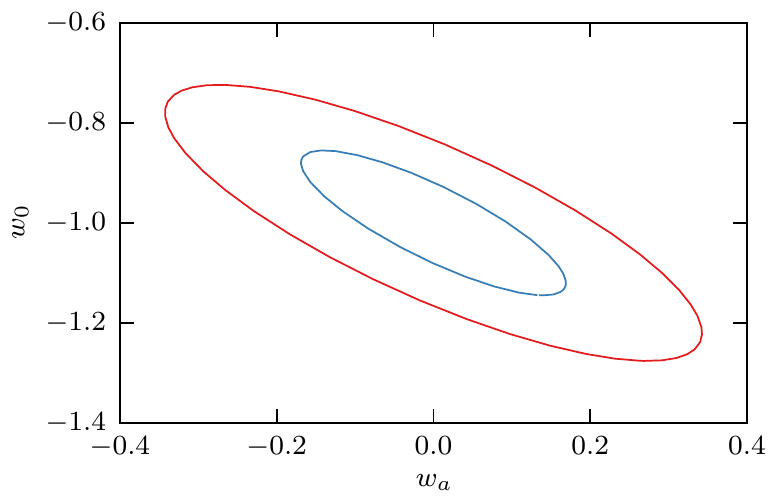}} 
\caption{Relative improvement in marginal errors of Dark Energy parameters, when IA systematics are included approximately in the shape analysis. Shape alone is shown in red (outer), and with size added in blue (inner). The FoM is increased by over a factor of 4.}
\label{fig:systematics_blob}
\end{figure}

Both size magnification and shear are subject to systematic errors.  In the latter case, a major source is intrinsic alignments \citep[IA; e.g.,][]{Heavens2000,Hirata2004}.  This can be converted into a statistical error if a flexible model is adopted \citep{Kirk2012}, where marginalising over the IA parameters increases the Dark Energy equation of state errors by a factor of 2-3.  Some of the lost FoM can be recovered with clustering information, leading to a degradation of Dark Energy errors by a factor of about 2 \citep{Joachimi2010,Kirk2012}. In the case of size, there is an anticorrelation between size and luminosity, which reduces the size magnification, because it is accompanied by flux magnification, which brings in less luminous and hence smaller galaxies into the sample.  This reduces the effect, dependent on the slope of the mean size-luminosity threshold. This depends on the mean size-luminosity relation, $\lambda \propto L^\beta$.  Estimates for $\beta$ vary  
%\cite{Simard2002} suggests $\beta= 0.29$, and 
%recent works vary, with 
slopes from 0.3 even up to unity \citep{Bernardi2012}.  Note that the effect is much diluted if the sample extends below $L*$, as the additional sources brought in by flux magnification are then a small proportion of the total, and the slope of the mean size-luminosity threshold is small.  At two magnitudes below $L*$ (the limit of Euclid at $z\simeq 1.8$) the effect is a few percent only\footnote{Note added for arXiV: This is true if no luminosity cuts are applied, when typically the dispersion will be larger.  It will be a significant effect if the analysis is done in luminosity bins.}.   An analogous effect to IA may exist in the form of size-size or size-density correlations.  Studies differ in their conclusions with current data \citep{Cooper2012,Papovich2012,Maltby2010,Rettura2010,Cimatti2012,Park2009}, and this will need careful study.  We will present a full study of size-shape weak lensing with systematics included in a later paper, but given that the size-magnitude effect is likely to be much smaller than the effects of IA, we expect the improvements presented here to be rather conservative.  To get a rough idea, increasing $\sigma_e$ by a factor of two approximates crudely the effect of  marginalising over IAs, by degrading the Dark Energy FoM by a similar factor.  This is illustrative only, as in reality the marginalization will lead to different contour shapes. In addition, the size signal is reduced by typically around a percent by size-luminosity correlations; equivalently we could increase the size noise by the same factor.  From the point-of-view of the improvements offered by size, we present conservative results by  increasing $\sigma_{\ln\lambda}$ by 10\% for the size-luminosity correlation. With these assumptions we find a relative improvement in the FoM by a large factor of 4.2 (Fig. \ref{fig:systematics_blob}).

\section{Conclusions}

In this paper we have shown that adding size measurement to cosmic shear analyses can lead to very significant improvements to the Dark Energy FoM of a weak gravitational lensing survey.  Ignoring systematics we find that the improvement is about 68\%, and we argue that much higher gains of even a factor 4 may be achievable when systematic effects are marginalised over, but this requires more detailed study.   The full gains can be achieved if the errors from size and shape are uncorrelated, and we have shown that for exponential profiles this can indeed be achieved, provided the square root of the area of the source is used as the measure of size.  We expect that for more general galaxy profiles and in the presence of PSF effects etc the correlation would be small, but non-zero.

\section*{acknowledgments}
We thank Benjamin Joachimi, Donnacha Kirk and Chris Duncan for discussions and for providing results which assisted this research.

\bibliographystyle{mn2e}
\bibliography{citations}

\begin{thebibliography}{28}
\expandafter\ifx\csname natexlab\endcsname\relax\def\natexlab#1{#1}\fi

\bibitem[{{Bartelmann} {et~al}\mbox{.}(1996){Bartelmann}, {Narayan}, {Seitz},
  \& {Schneider}}]{Bartelmann1996}
{Bartelmann} M., {Narayan} R., {Seitz} S., {Schneider} P., 1996, \apjl, 464,
  L115

\bibitem[{{Bernardi} {et~al}\mbox{.}(2012){Bernardi}, {Meert}, {Vikram},
  {Huertas-Company}, {Mei}, {Shankar}, \& {Sheth}}]{Bernardi2012}
{Bernardi} M., {Meert} A., {Vikram} V., {Huertas-Company} M., {Mei} S.,
  {Shankar} F., {Sheth} R.~K., 2012, ArXiv e-prints

\bibitem[{{Casaponsa} {et~al}\mbox{.}(2012){Casaponsa}, {Heavens}, {Kitching},
  {Miller}, {Bel{\'e}n Barreiro}, \&
  {Mart{\'{\i}}nez-Gonzalez}}]{Casaponsa2012}
{Casaponsa} B., {Heavens} A.~F., {Kitching} T.~D., {Miller} L., {Bel{\'e}n
  Barreiro} R., {Mart{\'{\i}}nez-Gonzalez} E., 2012, ArXiv e-prints

\bibitem[{{Castro} {et~al}\mbox{.}(2005){Castro}, {Heavens}, \&
  {Kitching}}]{Castro2005}
{Castro} P.~G., {Heavens} A.~F., {Kitching} T.~D., 2005, \prd, 72, 023516

\bibitem[{{Cimatti} {et~al}\mbox{.}(2012){Cimatti}, {Nipoti}, \&
  {Cassata}}]{Cimatti2012}
{Cimatti} A., {Nipoti} C., {Cassata} P., 2012, \mnras, 422, L62

\bibitem[{{Cooper} {et~al}\mbox{.}(2012){Cooper}, {Griffith}, {Newman}, {Coil},
  {Davis}, {Dutton}, {Faber}, {Guhathakurta}, {Koo}, {Lotz}, {Weiner},
  {Willmer}, \& {Yan}}]{Cooper2012}
{Cooper} M.~C. {et~al.}, 2012, \mnras, 419, 3018

\bibitem[{{Ferguson} {et~al}\mbox{.}(2004){Ferguson}, {Dickinson},
  {Giavalisco}, {Kretchmer}, {Ravindranath}, {Idzi}, {Taylor}, {Conselice},
  {Fall}, {Gardner}, {Livio}, {Madau}, {Moustakas}, {Papovich}, {Somerville},
  {Spinrad}, \& {Stern}}]{Ferguson2004}
{Ferguson} H.~C. {et~al.}, 2004, \apjl, 600, L107

\bibitem[{{Heavens} {et~al}\mbox{.}(2000){Heavens}, {Refregier}, \&
  {Heymans}}]{Heavens2000}
{Heavens} A., {Refregier} A., {Heymans} C., 2000, \mnras, 319, 649

\bibitem[{{Hildebrandt} {et~al}\mbox{.}(2009){Hildebrandt}, {van Waerbeke}, \&
  {Erben}}]{Hildebrandt2009}
{Hildebrandt} H., {van Waerbeke} L., {Erben} T., 2009, \aap, 507, 683

\bibitem[{{Hildebrandt} {et~al}\mbox{.}(2013){Hildebrandt}, {van Waerbeke},
  {Scott}, {B\'{e}thermin}, {Bock}, {Clements}, {Conley}, {Cooray}, {Dunlop},
  {Eales}, {Erben}, {Farrah}, {Franceschini}, {Glenn}, {Halpern},
  {et~al.}}]{Hildebrandt2013}
{Hildebrandt} H. {et~al.}, 2013, \mnras, 488

\bibitem[{{Hirata} \& {Seljak}(2004)}]{Hirata2004}
{Hirata} C.~M., {Seljak} U., 2004, \prd, 70, 063526

\bibitem[{{Hu}(2000)}]{Hu2000}
{Hu} W., 2000, \prd, 62, 043007

\bibitem[{{Hu}(2002)}]{Hu2002a}
{Hu} W., 2002, \prd, 65, 023003

\bibitem[{{Joachimi} \& {Bridle}(2010)}]{Joachimi2010}
{Joachimi} B., {Bridle} S.~L., 2010, \aap, 523, A1

\bibitem[{{Kirk} {et~al}\mbox{.}(2012){Kirk}, {Rassat}, {Host}, \&
  {Bridle}}]{Kirk2012}
{Kirk} D., {Rassat} A., {Host} O., {Bridle} S., 2012, \mnras, 424, 1647

\bibitem[{{Limber}(1954)}]{Limber1954}
{Limber} D.~N., 1954, \apj, 119, 655

\bibitem[{{Maltby} {et~al}\mbox{.}(2010){Maltby}, {Arag{\'o}n-Salamanca},
  {Gray}, {Barden}, {H{\"a}u{\ss}ler}, {Wolf}, {Peng}, {Jahnke}, {McIntosh},
  {B{\"o}hm}, \& {van Kampen}}]{Maltby2010}
{Maltby} D.~T. {et~al.}, 2010, \mnras, 402, 282

\bibitem[{{Miller} {et~al}\mbox{.}(2007){Miller}, {Kitching}, {Heymans},
  {Heavens}, \& {van Waerbeke}}]{Miller2007}
{Miller} L., {Kitching} T.~D., {Heymans} C., {Heavens} A.~F., {van Waerbeke}
  L., 2007, \mnras, 382, 315

\bibitem[{{Munshi} {et~al}\mbox{.}(2008){Munshi}, {Valageas}, {van Waerbeke},
  \& {Heavens}}]{Munshi2008}
{Munshi} D., {Valageas} P., {van Waerbeke} L., {Heavens} A., 2008, \physrep,
  462, 67

\bibitem[{{Papovich} {et~al}\mbox{.}(2012){Papovich}, {Bassett}, {Lotz}, {van
  der Wel}, {Tran}, {Finkelstein}, {Bell}, {Conselice}, {Dekel}, {Dunlop},
  {Guo}, \& et~al.}]{Papovich2012}
{Papovich} C. {et~al.}, 2012, \apj, 750, 93

\bibitem[{{Park} \& {Choi}(2009)}]{Park2009}
{Park} C., {Choi} Y.-Y., 2009, \apj, 691, 1828

\bibitem[{{Picinbono}(1996)}]{Picinbono1996}
{Picinbono} B., 1996, IEEE Transactions on Signal Processing, 44, 2637

\bibitem[{{Rettura} {et~al}\mbox{.}(2010){Rettura}, {Rosati}, {Nonino},
  {Fosbury}, {Gobat}, {Menci}, {Strazzullo}, {Mei}, {Demarco}, \&
  {Ford}}]{Rettura2010}
{Rettura} A. {et~al.}, 2010, \apj, 709, 512

\bibitem[{{Schmidt} {et~al}\mbox{.}(2012){Schmidt}, {Leauthaud}, {Massey},
  {Rhodes}, {George}, {Koekemoer}, {Finoguenov}, \& {Tanaka}}]{Schmidt2012}
{Schmidt} F., {Leauthaud} A., {Massey} R., {Rhodes} J., {George} M.~R.,
  {Koekemoer} A.~M., {Finoguenov} A., {Tanaka} M., 2012, \apjl, 744, L22

\bibitem[{{Shen} {et~al}\mbox{.}(2003){Shen}, {Mo}, {White}, {Blanton},
  {Kauffmann}, {Voges}, {Brinkmann}, \& {Csabai}}]{Shen2003}
{Shen} S., {Mo} H.~J., {White} S.~D.~M., {Blanton} M.~R., {Kauffmann} G.,
  {Voges} W., {Brinkmann} J., {Csabai} I., 2003, \mnras, 343, 978

\bibitem[{{Takada} \& {Jain}(2004)}]{Takada2004}
{Takada} M., {Jain} B., 2004, \mnras, 348, 897

\bibitem[{{Tegmark} {et~al}\mbox{.}(1997){Tegmark}, {Taylor}, \&
  {Heavens}}]{Tegmark1997}
{Tegmark} M., {Taylor} A.~N., {Heavens} A.~F., 1997, \apj, 480, 22

\bibitem[{{van Waerbeke}(2010)}]{VanWaerbeke2010}
{van Waerbeke} L., 2010, \mnras, 401, 2093

\end{thebibliography}

\end{document}